\documentclass[12pt]{article}
\usepackage{amsmath}
\usepackage{graphics}
\usepackage{epstopdf, color}

\begin{document}
{\bf \Large
\begin{center}
{Fisher's information for the position-dependent mass Schr\"odinger system}
\end{center}}
\vspace{2mm}
\begin{center}
{\large{\bf B. J. Falaye $^{a, }$}\footnote{E-mail:~ fbjames11@physicist.net},  {\large{\bf F. A. Serrano $^{b,}$}\footnote{E-mail:~ univeresime@hotmail.com}, and {\large{\bf Shi-Hai Dong $^{c,}$}\footnote{Corresponding author: E-mail:~ dongsh2@yahoo.com; Tel:+52-55-57296000 ext.52522.}}}}
\end{center}
{\small
\begin{center}
{\it $^\textbf{a}$ESFM, Instituto Polit\'{e}cnico Nacional, UPALM, M\'{e}xico D. F. 07738, M\'{e}xico}\\
{\it $^\textbf{b}$Escuela Superior de Ingenier\'{\i}a Mec\'{a}nica y El\'{e}ctrica UPC, Instituto Polit\'{e}cnico Nacional,
Av. Santa Ana 1000, M\'{e}xico, D.F. 04430, M\'{e}xico}\\
{\it $^\textbf{c}$CIDETEC, Instituto Polit\'{e}cnico Nacional, UPALM, M\'{e}xico D. F. 07700, M\'{e}xico}
\end{center}}
\begin{center}
{\bf Phys. Lett. A (2015) doi:10.1016/j.physleta.2015.09.029}
\end{center}
\noindent
\begin{abstract}
\noindent
This study presents the Fisher information for the position-dependent mass Schr\"odinger equation with hyperbolical potential {$V(x)=-V_0{\rm csch}^2(ax)$}. The analysis of the quantum-mechanical probability for the ground and exited states $(n=0, 1, 2)$ has been obtained via the Fisher's information. This controls both chemical and physical properties of some molecular systems. The Fisher information is considered only for $x>0$ due to the singular point at $x=0$. We found that Fisher-information-based uncertainty relation and the Cramer-Rao inequality holds. Some relevant numerical results are presented. The results presented shows that the Cramer-Rao and the Heisenberg products in both spaces provide a natural measure for anharmonicity of {$-V_0{\rm csch}^2(ax)$}.
\end{abstract}
\noindent
{\bf PACs No:} 	 03.65.-w, 03.65. Ge, 03.67.-a.

\section{Introduction}
In recent years, there has been a great interest in studying information theoretic measures for different quantum systems. This is due to the fact that information theory of quantum-mechanical systems is related to the modern quantum communications, computation and the density functional methods \cite{A1}. According to the density functional theory (DFT) initiated by Hohenberg and Kohn \cite{A2}, the one-particle position and momentum probability densities are the basic elements for describing the physical and chemical properties of some molecular systems. The quantum information theory plays an important role in the measure of uncertainty and other quantum parameters of the system. The main measures of quantum information are the Shannon entropy \cite{A3} and Fisher information \cite{A4}. They are function of a characteristic probability density. They are traditionally used in engineering, physics, applied mathematics, condensed physics, chemical and other related areas.

The Fisher information was introduced by Fisher as a measure of intrinsic accuracy in statistical estimation theory but its basic properties are not completely well known yet, despite its early origin in 1925 [5]. The importance of this was noticed by Sears et al. \cite{A6}. The authors found that the quantum mechanical kinetic energy can be considered as a measure of the information distribution. Fisher information has been very useful and has been applied in different areas. For example using the principle of minimum Fisher information \cite{A7}, one can obtain the equations of non-relativistic quantum mechanics \cite{A8}, the time-independent Kohn--Sham equations and the time-dependent Euler equation of DFT \cite{A9}. Its local character is the main difference with respect to Shannon information which is global information measure. It is defined as the expectation value of the logarithmic gradient of density or as the gradient functional of density. So the Fisher information is given by \cite{A5}

\noindent
\begin{equation}
\label{EQ1}
I_F =\int _{-\infty}^{\infty}{\rho }(x)\left[\frac{d}{dx}\ln{\rho}(x)\right]^2 {d}{x}=\int_{a}^{b}\frac{\left[{\rho}'(x)\right]^{2} }{\rho(x)}dx .
\end{equation}
If the probability density is defined as $\rho_n(x)=\left|\psi_n(x)\right|^{2}$, then
\begin{equation}
\label{EQ2}
{I}_F =\int _{a}^{b}\left|\psi(x)\right|^2\left[\frac{d}{dx}\ln\left|\psi(x)\right|^2\right]^2{d}{x}=4\int_{a}^{b}\left[\psi'( x)\right]^2{d}{x},
\end{equation}
which is not totally independent. There is an inequality which involves Fisher information and variance ${V}=\left\langle{x}^{2} \right\rangle-\left\langle{x}\right\rangle^{2}$. It is called the Cramer-Rao uncertainty relation:
\begin{equation}
\label{EQ3}
I_F \cdot{V}\ge{1}.
\end{equation}
The Fisher information is a derivative functional of the density, so that it is very sensitive to local rearrangements of ${\it \rho }_{{\it n}} ({\it x})$. In this paper we present the Fisher information of the position-dependent mass Schr\"odinger equation with hyperbolical potential.

The study of the Schr\"odinger equation with a position-dependent mass (PDM) is a very useful model of interest since the early days of solid state physics and in many applied branches of quantum physics such as condensed matter physics, material science, nuclear physics etc. Special application of principal concept of PDM is found in the investigation of electronic properties of semiconductors, quantum dots and wells, etc \cite{A10}.

The rest part of this work is organized as follows: In the section 2, we give a brief review of the position-dependent mass Schr\"odinger equation. A particular case of the hyperbolic potential is presented. In section 3, we first present the normalized wave function in position space and then calculate the Fisher Information $I_F $, the Heisenberg uncertainty product and Cramer-Rao product of hyperbolic cosecant potential for various values of potential parameter $a$ and for few states $n=0, 1, 2$. Finally, we give some concluding remarks in section 4.

\section{Calculation of the wave functions}
The Schr\"odinger equation with the position-dependent mass for an arbitrary potential $V(x)$, can be expressed as \cite{A10, A11, A12}
\begin{equation}
\label{EQ4}
\nabla_x\left(\frac{1}{m(x)}\nabla_x\psi(x)\right)+2m_0\left[E-V(x)\right]\psi(x)=0,
\end{equation}
where $E$ is the energy spectrum and solitonic smooth effective mass distribution {$(m(x))$} is taken as $m(x)= m_{0} (x)\mbox{sech}^{2}(ax)$, which has been used widely in condensed matter and low-energy nuclear physics. Taking $\psi(x)=\cosh^{\tau}(ax)\mathcal{F}(ax)$ and then substitute it into equation \eqref{EQ4}, we have
\begin{eqnarray}
\label{EQ5}
&&\mathcal{F}''(ax)+2a(1+\tau)\tanh(ax)\mathcal{F}'(x)+\frac{\mbox{sech}^2(ax)}{2}\left\{a^2\tau(\tau+2)\cosh(2ax)\right.\nonumber\\
&&\left.+\left[4m_0\left(E-V(x)\right)-a^2\tau^2\right]\right\}\mathcal{F}(ax)=0.
\end{eqnarray}
Further substitution of $\delta =2m_0/a^2$ and $\gamma=ax$ into equation (\ref{EQ5}) gives
\begin{equation}
\label{EQ6}
\mathcal{F}''(\gamma)+2(1+\tau)\tanh(\gamma)\mathcal{F}'(\gamma)+\left\{\tau(\tau+2)\tanh^{2}(\gamma)+\left[\tau+\sigma \left(E-V(y)\right)\right]\mbox{sech}^{2}(\gamma)\right\}\mathcal{F}(\gamma)=0.
\end{equation}
Considering a new relation of the form $\mbox{sech}(\gamma)=\cos(z)$ and $\tanh(\gamma)=\sin(z)$, which transform the boundary condition of the wave function from $(-\infty, \infty)$ to {$\left(-\pi/2, \pi/2\right)$}  and taking $\tau =-1/2$, then the above equation (\ref{EQ6}) can be simplified further as
\begin{equation}
\label{EQ7}
-\mathcal{F}''(z)+\mathcal{V}(z)\mathcal{F}(z)=\varepsilon\mathcal{F}(z)\ \ \mbox{with}\ \ \mathcal{V}(z)=\frac{3}{4}\tan^{2}(z)+\sigma V(z)+\frac{1}{2}, \ \ \varepsilon =\delta{E}.
\end{equation}
In recent study \cite{A14}, the Shannon entropy for the position-dependent Schr\"odinger equation for a particle with a nonuniform solitonic mass density is evaluated in the case of a trivial null potential. It was found that the negative Shannon entropy exists for the probability densities that are highly localized. In this work, we consider a special squared hyperbolic cosecant potential  $V(ax)= -V_0\mbox{csch}^2(ax)$ and then analyze its quantum-mechanical probability cloud for the ground and excited states by means of local (Fisher's information) information-theoretic measure. Now, substituting this potential into equation \eqref{EQ7} and recalling the relation $\mbox{sech}(\gamma)=\cos(z)$, one has {$\mathcal{V}(z)=3\tan^{2}(z)/4-\delta V_0\cot^{2}(z)+1/2$}. It is interesting to note that this family of potential represents different potentials in z space. For example, for $\mathcal{V}_0=\delta V_0 >0$, they look like infinitely deep funnels and behave like the potential $1/{x}$, while for $\mathcal{V}_{0}<0$ they become infinite double-wells and if $\mathcal{V}_0 =0$ they become the infinite single-well.

In order to obtain exact solution to this system, we take the following wave function ansatz:
\begin{equation}
\label{EQ8}
\mathcal{F}(z)=\sin^{\mu}(z)\cos^{\nu}(z)\mathcal{G}(z),
\end{equation}
where the parameters ${\it \mu }$ and ${\it \nu }$ are calculated by considering the behaviors of the wave functions at $z\sim{0}$ and {${z}\sim \pi/2$ as $\mu=1/2+\sqrt{1-4\mathcal{V}_{0}}/2$ and $\nu = 3/2$} respectively. The function $\mathcal{G}(z)$ satisfies the following differential equation
\begin{equation}
\label{EQ9}
\mathcal{G}''(z)+2\left[{\mu }\cot(z)-\nu\cot(z)\right]\mathcal{G}'(z)+\left(\varepsilon -\mu-\nu-2\mu\nu\right)\mathcal{G}(z)=0.
\end{equation}
Using a change of variable $\xi=\sin^{2}(z)$, the equation is transformed to
\begin{equation} \label{EQ10}
\xi(1-\xi)\mathcal{G}''(\xi)+\left[\mu+\frac{1}{2} -(1+\mu+\nu)\xi\right]\mathcal{G}'(\xi)+\frac{1}{4}\left(\varepsilon-\frac{1}{2}-\mu -\nu-2\mu\nu\right)\mathcal{G}(\xi)=0,
\end{equation}
whose solution is given by hypergeometric function $\ _2F_{1}(a, b; c; \xi)$ with the parameters
\begin{equation}
\label{EQ11}
a=\frac{\mu}{2}+\frac{\nu}{2}-\frac{1}{2}\sqrt{\varepsilon+\frac{1}{4}-\mathcal{V}_0}, \ \ \ b=\frac{\mu}{2}+\frac{\nu }{2}+\frac{1}{2} \sqrt{\varepsilon +\frac{1}{4} -\mathcal{V}_0}, \ \ \ c=\mu+\frac{1}{2}.
\end{equation}
Based on the quantum condition ${\it a}={\it -n}$, which makes the hypergeometric functions terminate to a polynomial, thus we  might obtain the energy levels as
\begin{equation}
\label{EQ12}
\varepsilon _{n} =4\left(n+1\right)\left[n+1+\sqrt{\frac{1}{4}-\mathcal{V}_0}\right], \ \ \ \mathcal{V}_0 \le\frac{1}{4}, \ \ \ n=0, 1, 2, ....
\end{equation}
The corresponding wave functions are given by
\begin{eqnarray}
\label{EQ13}
\psi(x)&=&N_n\tanh^{\mu}(ax)\mbox{sech}^{2}(ax)\ _2F_1\left(-n, \mu+\nu+n; \mu+\frac{1}{2}; \tanh^{2}(ax)\right) \nonumber\\
\psi(\varrho)&=&{\widetilde{N_n}\left(\frac{1-\varrho}{2}\right)^{\frac{\mu}{2}}\left(\frac{1+\varrho}{2}\right)P_n^{\left({\mu-\frac{1}{2}}, \nu-\frac{1}{2}\right)}(\varrho), ~~~~~ \varrho=1-2\tanh^2(ax)},
\end{eqnarray}
where the normalization factor is obtained as
\begin{equation}
\label{EQ14}
\begin{array}{l}
\widetilde{N_{n}}=\displaystyle\sqrt{\frac{2a(\mu+3/2+2n)\Gamma(\mu+3/2+n)\Gamma(\mu+1/2+n)}{(1+n)\Gamma(\mu+1/2)^2[(\mu+1/2)_{n}]^2}}\\[3mm]
=\displaystyle\sqrt{\frac{2a\left(\mu+\frac{3}{2}+2n\right)\left(\mu+\frac{1}{2}+n\right)}{n+1}},
\end{array}
\end{equation}
where we have used the Pochhammer symbol $(x)_{n}=\Gamma(x+n)/\Gamma(x)$  with $\Gamma(n+1)=n\Gamma(n)$ and formula 739.1 of ref \cite{MM}. Since there exists a singular point at $x=0$ for the potential, then we will consider the wave functions at the interval $x>0$.

\section{Fisher information and uncertainty principle}
Fisher information provides the main theoretic tool of the extreme physical information principle and a general variational principle which allows one to derive various fundamental equations of physics. If the corresponding bounds on the degree to which members of a family of quantum states can be distinguished by measurement, then quantum generalizations of Fisher information may be given. In ref. \cite{A15}, the Fisher information of a quantum observable is shown to be proportional to both the difference of a quantum and a classical variance, thus providing a measure of nonclassicality and the rate of entropy increase under Gaussian diffusion, thus providing a measure of robustness. Dehesa et al. \cite{A16} obtained the spreading of the quantum-mechanical probability distribution density of D-dimensional hydrogenic orbitals via the local information-theoretic quantity of Fisher in both position and momentum spaces. Other relevant works includes (\cite{A17, A18, A19} and refs. therein)

However, Fisher information is yet to be evaluated analytically for the position-dependent mass Schr\"odinger system. To achieve this goal, we utilize the wave function obtained for this system to find the probability and then analyze the quantum-mechanical probability by the means of Fisher's information. We consider only $x>0$ due to the singular point at $x=0$ and we found that Fisher information based uncertainty relation and the Cramer-Rao inequality holds.

From equation (\ref{EQ2})
\begin{eqnarray}
\label{EQ15}
I_F&=&4\int _{0}^{\infty}\left[\psi'( x)\right]^2{d}{x}=16a\int_{-1}^{1}\left(\frac{1+\varrho}{2}\right) \left(\frac{1-\varrho}{2}\right)^{1/2}\left[\psi'(\varrho)\right]^2{d}{\varrho}\nonumber\\
   &=&\widetilde{N_{n}}^2\int_{-1}^{1}\left\{\left[(3+2n+2\mu)(\varrho^2-1)P_{n-1}^{\left(\mu+\frac{1}{2}, 2\right)}(\varrho)+2\left[2(\varrho-1)+\mu(\varrho+1)\right]P_{n}^{\left(\mu-\frac{1}{2}, 1\right)}(\varrho)\right]^2\right.\nonumber\\
&&\times\left.\frac{a}{16}\left(\frac{1+\varrho}{2}\right)\left(\frac{1-\varrho}{2}\right)^{\mu-\frac{3}{2}}\right\}{d}{\varrho}, ~~~n>1.
\end{eqnarray}
To analytically evaluate the integral in equation equation (\ref{EQ15}) for {$n$}-state is very difficult. For convenience we only study a few low-lying normalized states {$n = 0, 1, 2, 3$} to calculate the Fisher information.
\begin{itemize}
\item For $n=0$, we have
\begin{eqnarray}
\label{EQ16}
I_F^{(0)}&=&\frac{32a\left(2\mu^2+4\mu-1\right)}{(2\mu-1)(2\mu+1)(2\mu+3)(2\mu+5)}\widetilde{N_{0}}^2.\ \ \ \mbox{ Taking $\mathcal{V}_0 = 2^{-5}$, then}\nonumber\\
I_F^{(0)}&=&\frac{128a\left(15869-228\sqrt{14}\right)}{1190035}\widetilde{N_{0}}^2.
\end{eqnarray}
\item Similarly, for $n=1$
\begin{eqnarray}
I_F^{(1)}&=&\frac{128a\widetilde{N_{1}}^2\left(\mu+\frac{1}{2}\right)_1^{-1}(-33+68\mu+52\mu^2+8\mu^3)}{(2\mu-1)(2\mu+3)(2\mu+5)(2\mu+7)(2\mu+9)}\ \ \ \mbox{Taking $\mathcal{V}_0 = 2^{-5}$, then}\nonumber\\
         &=&\frac{1024a\widetilde{N_{1}}^2 \left(366010057-7493496 \sqrt{14}\right)}{95313473255\left[\frac{1}{8}\left(4+\sqrt{14}\right)+\frac{1}{2}\right]}.
\end{eqnarray}
\item For $n=2$
\begin{equation}
I_F^{(2)}=-\frac{2304a\widetilde{N_{2}}^2\left(29629036 \sqrt{14}-2060523213\right)}{398279448385\left[\frac{1}{8} \left(4+\sqrt{14}\right)+\frac{1}{2}\right]\left[\frac{1}{8}\left(4+\sqrt{14}\right)+\frac{3}{2}\right]}.
\end{equation}
\item For $n=3$
\begin{equation}
{I_F^{(3)}}=-\frac{49152a\widetilde{N_{3}}^2\left(8867016 \sqrt{14}-809394467\right)}{831817071085 \left[\frac{1}{8} \left(4+\sqrt{14}\right)+\frac{1}{2}\right] \left[\frac{1}{8} \left(4+\sqrt{14}\right)+\frac{3}{2}\right]\left[\frac{1}{8} \left(4+\sqrt{14}\right)+\frac{5}{2}\right]}.
\end{equation}
\end{itemize}
We give a useful remark on the choice of $\mathcal{V}_0$. In terms of the parameter $\mu$ as given below Eq.(\ref{EQ8}) and the energy $E_n$ given in (\ref{EQ12}), we have $\mathcal{V}_0\leq 1/4$. For convenience, in the calculation we take $\mathcal{V}_0=1/32$. We proceed further to study the  uncertainty relations. Uncertainty relations  form the basic properties of quantum mechanics. Particularly, the Heisenberg uncertainty principle which state that the product the uncertainties in position and momentum, can be expressed in terms of Planck's constant, i.e. $\Delta(x)\Delta(p)\ge\frac{\hbar}{2}$.In order to obtain this relation, it is required to calculate the expectation values of $\left\langle x\right\rangle_n$, $\left\langle p\right\rangle_n$, $\left\langle x^2\right\rangle_n$ and $\left\langle p^2\right\rangle_n$ as follows:
\begin{equation}\left\langle x\right\rangle_n=\int_0^\infty \psi(x)x\psi(x)dx=\widetilde{N}_{n}^2\int_0^{1}\frac{\left[\tanh^{-1}\left(\sqrt{y}\right)\right]}{2a^2y^{\frac{1}{2}-\mu}}(1-y)\left[P_n^{\left({\mu-\frac{1}{2}}, 1\right)}(1-2y)\right]^2dy.
\label{E20}
\end{equation}
\begin{equation}\left\langle x^2\right\rangle_n=\int_0^\infty \psi(x)x^2\psi(x)dx=\widetilde{N}_{n}^2\int_0^{1}\frac{\left[\tanh^{-1}\left(\sqrt{y}\right)\right]^2}{2a^3y^{\frac{1}{2}-\mu}}(1-y)\left[P_n^{\left({\mu-\frac{1}{2}}, 1\right)}(1-2y)\right]^2dy.
\label{E21}
\end{equation}
\begin{equation}
\left\langle p\right\rangle_n=0, \ \ \ \left\langle p^2\right\rangle_n=\int_0^\infty\psi(x)\left(-\frac{d^2}{dx^2}\right)\psi(x)dx, \ \ \ \ \ \ \ y={\rm tanh}^2(ax).
\label{E22}
\end{equation}

\begin{table}[!t]
\begin{center}
{\scriptsize
\caption{\normalsize Numerical results for the uncertainty relation and Fisher information measure for the position-dependent mass Schr\"odinger system}
\vspace{6mm}{
\begin{tabular}{cccccccccc}\hline\hline\\
{$n$}&	{$a$}&$\left\langle x^2\right\rangle$&$\left\langle x\right\rangle$&$\Delta(x)$&$\left\langle p^2\right\rangle$&$\Delta(p)$&$\Delta(x)\Delta(p)$&$I_\rho$  &	$I_\gamma$\\[1ex]\hline\hline
0&	1&	0.808403&	0.809678&	0.390927&	2.92486&	1.71022&	0.66857&	3.23361&	11.6994\\[1ex]
 &	2&	0.202101&	0.404839&	0.195463&	11.6994&	3.42044&	0.66857&	0.80840&	46.7977\\[1ex]	
 &	4&	0.050525&	0.202420&	0.097732&	46.7977&	6.84089&	0.66857&	0.20210&	187.191\\[1ex]
1&	1&	1.302240&	0.958246&	0.619683&	10.0079&	3.16353&	1.96039&	5.20897&	40.0318\\[1ex]
 &	2&	0.325560&	0.479123&	0.309841&	40.0318&	6.32707&	1.96039&	1.30224&	160.127\\[1ex]
 &	4&	0.081390&	0.239561&	0.154921&	160.127&	12.6541&	1.96039&	0.32556&	640.508\\[1ex]
2&	1&	1.551850&	1.019370&	0.716054&	21.0797&	4.59127&	3.28760&	6.20740&	84.3190\\[1ex]
 &	2&	0.387963&	0.509686&	0.358027&	84.3190&	9.18254&	3.28760&	1.55185&	337.276\\[1ex]	
 &	4&	0.096991&	0.254843&	0.179013&	337.276&	18.3651&	3.28760&	0.38796&	1349.10\\[1ex]	

\hline\hline
\end{tabular}\label{tab2}}
\vspace*{-1pt}}
\end{center}
\end{table}

Let us derive some integrals for special cases $n=0, 1, 2$. When $n=0$, we have
\begin{equation}\label{}
\left\langle x\right\rangle_0=\frac{\left(\mu +\frac{1}{2}\right) \left(\mu +\frac{3}{2}\right) \left(2 (\mu +1) H_{\mu }+\mu  (\log (16)-2)-1+\log (16)\right)}{a (\mu +1) (2 \mu +1) (2 \mu +3)};
\end{equation}
\begin{equation}\label{}
\left\langle x\right\rangle_1=\frac{8 (\mu +1) (\mu +2) (\mu +3) H_{\mu }-12 \mu ^3-52 \mu ^2-43 \mu +16 (\mu +1) (\mu +2) (\mu +3) \log (2)+18}{16 a (\mu +1) (\mu +2) (\mu +3)}
\end{equation}

\begin{equation}\label{}
\begin{array}{l}
\left\langle x\right\rangle_2=\displaystyle\frac{1}{192 a (\mu +1) (\mu +2) (\mu +3) (\mu +4) (\mu +5)
}\Big\{-176 \mu ^5-2208 \mu ^4-9488 \mu ^3-15192 \mu ^2\\
-3467 \mu +96 \gamma  (\mu +1) (\mu +2) (\mu +3) (\mu +4) (\mu +5)\\
+192 (\mu +1) (\mu +2) (\mu +3) (\mu +4) (\mu +5) \log (2)\\
+96 (\mu +1) (\mu +2) (\mu +3) (\mu +4) (\mu +5) \psi^{(0)}(\mu +1)+6735\Big\}
\end{array}
\end{equation} where $H_{\mu}$ and $\psi^{0}$ denote the Harmonic number and digamma function, respectively. For a given ${\mathcal{V}_0}=1/32$, one has $\mu=0.967707$. Thus above average values can be simplified as $a\left\langle x\right\rangle_0=0.809678, a\left\langle x\right\rangle_1=0.958246, a\left\langle x\right\rangle_2=1.01937$. In a similar way, we are able to obtain the following average values  $a^2\left\langle x^2\right\rangle_0=0.808403, a^2\left\langle x\right\rangle_1=1.30224, a^2\left\langle x^2\right\rangle_2=1.55185$. The numerical results are presented in Table 1. Furthermore, we can obtain the Fisher information for the expectation values of the position and momentum via $I_\rho=4\left\langle x^2\right\rangle$ and $I_\gamma=4\left\langle p^2\right\rangle$. We find that the $I_{\rho}$ decreases with the increasing width of the mass barrier $a$ while $I_{\gamma}$ increases with it. Neverthelss, the relation $I_\rho\cdot I_\gamma\ge D^2$ holds for Cramer-Rao uncertainty products \cite{A20}. The results shows that the Heisenberg uncertainty principle holds for various values of parameter $a$. The results in Table 1 agree with this relation.

\section{Concluding remarks}
The PDM Schr\"odinger equation for a particle with a nonuniform solitonic mass density is evaluated in the case of a non trivial potential.  We consider the special squared hyperbolic cosecant potential as a model. Firstly, we find the wave function with the corresponding normalization factor by considering only $x>0$ due to the singular point at $x=0$. The Fisher information of the quantum system have been studied. This is a local theoretic quantity which measures the spreading of the quantum mechanical probability density. On the other hand Fisher information describes the concentration of the density around its nodes, providing a measure of the oscillatory character of the corresponding wave function. We found that the Fisher-information-based uncertainty relation and the Cramer-Rao inequality holds. The validity Heisenberg Uncertainty principle for various values of potential parameter $a$ has been investigated for this potential model. The results presented shows that the Cramer-Rao and the Heisenberg products in both spaces provide a natural measure of the deviation of {$-V_0{\rm csch}^2(ax)$} from being a harmonic oscillator.

\vspace{6mm} {\Large \bf Acknowledgments}: We would like to thank the kind referees for positive and invaluable suggestions which have improved the manuscript greatly. 
This work was supported partially by project 20150964-SIP-IPN, COFAA-IPN, Mexico.

\end{document}